\font\lbf=cmbx10 scaled\magstep2
\font\sbf=cmr10 
\def\bs{\bigskip}
\def\ms{\medskip}

\def\ni{\noindent}
\def\cl{\centerline}
\def\title#1{\cl{\lbf #1}\ms}

\def\dtitle#1{\bs\cl{\sbf #1}\par\nobreak\ms}
\def\stitle#1{\bs{\ni \bf #1}\par\nobreak\ms}
\def\ref#1#2#3#4#5{#1\ #2\ {\it#3}{~~\bf#4}~#5\par}
\def\ns{\kern-.33333em}
\magnification=\magstep1

\title{Cosmological PPN Formalism and}
\title{Non-Machian Gravitational Theories}
\dtitle{G. Dautcourt 
\footnote{$^1$}{Humboldt-Universit\"at zu Berlin, 
Institut f\"ur Physik, D-10115 Berlin, Germany }}

\bs \bs \ni {\it received~~~~~~~~~~~} 
\vskip 5pt
\hrule 
\vskip 10pt

\vskip 10pt
\hrule
\vskip 5pt
\ms\ni
\stitle{1. INTRODUCTION}\ni
There appears to be little agreement what 
Mach's principle actually is -  this was my chief impression from a recent 
conference on Mach's principle at T\"ubingen (Germany). Should we relate it 
to the requirement that the inertial mass of a body is determined by the
remaining bodies in the Universe? Or should we follow one of Barbours
suggestions (Barbour 1995), perhaps that theories must be formulated in 
terms of relative quantities to become Machian, since we observe relative
quantities (positions, velocities) only?
Mach himself seems to be of little help in this respect since for almost 
every formulation of a Machian principle one may find an appropriate citation 
in his publications.

Another item from Mach's General Store (Brill 1995) is the rather vague 
statement that distant parts of the Universe should have some influence 
on the form of local laws of physics. In a non-Machian gravitational
theory a local experiment should therefore be {\it independent} 
of the cosmological environment. In particular, no fields of cosmological 
origin should have an influence on the local motion of matter, "local" 
being taken to denote a sufficiently small space region.
The purpose of this note is to show that - within a post-Newtonian
approximation - metric theories of gravity 
may be divided according to the requirement of {\it time-dependent 
potentials} for a consistent local description of cosmological models. 
This kind of dependence on cosmological boundary conditions is used 
to define Machian and non-Machian theories. 

To deal with a fairly 
general class of metric gravitational theories, the Parametrized 
Post-Newtonian (PPN-) formalism by Kenneth Nordtvedt and Clifford M. Will 
will be used. The notation is taken from Misner et al. (1973), otherwise 
I follow Will's standard book (Will 1993).

The PPN formalism was clearly not designed to deal with cosmology.
On the other hand, one has to face the fact that a Newtonian 
cosmology was developed decades ago 
(see, e.g., Heckmann and Schuecking 1955, 1956, 1959, Trautman 1966), with 
results very similar to those of General Relativity. The relation of this
Newtonian cosmology to the Friedman models remained unclear, 
however.  One expects that a Newtonian cosmology 
should emerge as a first approximation of the general-relativistic 
theory. Then one may ask for a post-Newtonian cosmology as a second-order 
approximation. Notice the Newtonian cosmology as discussed here is based 
on a Lorentzian
manifold, and hence quite different from the Newton-Cartan formulation of
a transition to Newton's gravity (for the latter see, e.g., Trautman 1956, 
Ehlers 1981, Lottermoser 1988, Dautcourt 1990, and references therein).

Somewhat surprisingly, the PPN formalism {\it as it stands } is already 
able to some extent to deal with cosmological problems, as will be shown
subsequently.      
The keys are (i) to use differential relations 
to avoid the Minkowskian boundary conditions at spatial infinity in the 
usual integral formulation of the PPN formalism, and (ii) to take the 
ratio distance (measured from the center of a selected spatial region)
to  Hubble distance $c/H_0$ as the expansion parameter. This leads to
a simple local description of cosmological models, which should be valid 
for {\it small} values of the expansion parameter.

Section 2 summarizes 
basic aspects of Newtonian cosmology. In Section 3 
PPN relations as well as the divergence freedom of the matter tensor are 
used to construct locally isotropic cosmological solutions beyond the
Newtonian approximation. 

Restricting to homogeneous models in Section 4, an inconsistency is found 
in generic metric theories (with the exception of the non-Machian class) 
at the post-Newtonian level. 
We suggest that the discrepancy might be caused by the neglection of 
time-dependent components in the Newtonian and post-Newtonian potentials 
at the origin of the PPN coordinate system.

Section 5 illuminates the problem from a different point of view:
Starting from a comoving cosmological coordinate system, in which the
cosmological fluid with a divergence-free matter tensor is homogeneous
and expands isotropically, a transformation into the standard PPN 
coordinate system produces generically time-dependent local potentials. 
Again the potentials are absent in a small subset of theories, formed 
by a non-Machian class of gravitational theories.

\stitle{2. NEWTONIAN COSMOLOGY}
\ni
The PPN formalism assumes the existence of a coordinate system where the 
metric tensor can be written as 
$$\eqalignno {g_{00} &= -1 + 2U -2\beta U^2 + 4\Psi -\zeta {\cal A} 
+ O(\epsilon^6), &(1)\cr
g_{0i} &= -{7\over2}\Delta_1 V_i - {1\over2} \Delta_2 W_i + O(\epsilon^5), 
&(2)\cr g_{ik} &= \delta_{ik}(1 + 2\gamma U) + O(\epsilon^4). &(3)\cr} $$

The space-time functions $U, \Psi,{\cal A}, V_i, W_i$ are defined in terms of 
the matter density, the pressure $p$ and velocity components $v_i({\bf x},t)$ 
of the fluid: 
$$\eqalignno{U({\bf x},t) &=\int d{\bf x'}\rho({\bf x'},t)/\mid \bf x 
-\bf x' \mid,  &(4) \cr
\Psi({\bf x},t)&=\int d{\bf x'}\rho({\bf x'},t) \phi({\bf x'},t)
    /\mid \bf x -\bf x' \mid, &(5) \cr
V_i({\bf x},t) &= \int d{\bf x'}\rho({\bf x'},t) v_i({\bf x'},t)
    /\mid{\bf x} - {\bf x'} \mid,  &(6) \cr
W_i({\bf x},t) &=  \int d{\bf x'}\rho({\bf x'},t)[({\bf x-x}'){\bf v(x'},t)] 
    (x^i-x'^i) /\mid \bf x -\bf x' \mid^3, &(7) \cr 
{\cal A}({\bf x},t)&= \int d{\bf x'}\rho({\bf x'},t)[({\bf x 
    - x'}){\bf v(x'},t)]^2/\mid \bf x -\bf x' \mid^3,   &(8) \cr}$$
where
$$\phi({\bf x},t) = \beta_1v^2 + \beta_2U +{3\over2}\beta_4p/\rho_0. \eqno(9)
$$

As it stands, the PPN formalism appears not to be applicable to cosmology: 
If a homogeneous matter density $\rho(t)$ is introduced, the Newtonian 
potential $U$ as calculated from (4) diverges - this is the well-known
Seeliger-Neumann paradoxon, seen to be present also in a higher-order
correction to the Newtonian potential.  
A simple  way to overcome this difficulty was proposed by Heckmann and 
Sch\"ucking (1959), see also  Trautman (1966): In agreement with the 
equivalence principle, the Newtonian potential and its first derivatives are 
considered as quantities subject to changes induced by coordinate 
transformations beyond the Galileo group. The second derivatives only, which 
are determined by the generating mass density, should have a physical meaning.
Thus the function
$$ U = -{2\over3}\pi G \rho(t) r^2, \eqno(10) $$
was taken as the potential in Newtonian cosmology.
It is evidently a solution of the Poisson equation for $U$,
$$ \Delta U = -4\pi\rho, \eqno(11) $$
generated by  homogeneous matter density. The divergence of $U$ for large 
$r$ can be considered as coordinate singularity, now caused by the use of a 
local (non-cosmological) coordinate system. The singularity vanishes, if a 
transformation to cosmological (comoving) coordinates is performed, 
see below Eq.(19).

The post-Newtonian  approximation as applied to objects in the Solar system 
uses two expansion parameters $\epsilon= \sqrt{\mid U_m \mid}$ and $v/c$, 
where $U_m$ is the maximal value of the Newtonian potential and $v$ is the
velocity of bodies moving under the influence of gravity. For the bookkeeping 
of the different approximations both parameters are usually considered as of
equal order. This is justified also in the case considered here: 
The cosmological potential (10) gives 
$\epsilon \approx r\sqrt{2\pi\rho_0G/3} \approx r/L$, where $L$ is a typical
Hubble distance. The radial velocity of 
galaxies is of equal order as is evident from Hubbles redshift-distance 
relation. As dimensionless expansion parameter $\epsilon$ one may therefore 
take the ratio distance to Hubble distance.

To Newtonian approximation the metric tensor is given by          
$$\eqalignno{g_{00} &=-1 +2U + o(\epsilon^4), &(12) \cr 
g_{0i} &= o(\epsilon^3), &(13)\cr
g_{ik} &= \delta_{ik} + o(\epsilon^2), &(14)\cr }$$
and the dynamical equations reduce in the case of dust matter, which will 
be assumed for simplicity throughout this note, to 
$$ \eqalignno{\partial\rho/\partial t +\partial (\rho v^i)/\partial x^i 
&=0, &(15)\cr \rho(\dot v_i +v_{i,k}v_k) &= \rho U_{,i}, &(16) \cr }$$
where the potential $U$ is given by the solution (10) of the Poisson 
equation (11). For an isotropic expansion with a scale factor $R(t)$ 
$$ v^i = x^i f_0(t) = x^i\dot R/R,\eqno(17) $$
(15) and (16) together with the potential (10) lead to matter conservation 
and  to the Friedman equation:
$$ R^3\rho = const, ~~\dot R^2 = 2G{\cal M}/R +const.\eqno(18) $$
This similarity between Newtonian and general-relativistic 
cosmology has sometimes been taken as evidence that Newtonian gravity might 
have a wide range of applicability in cosmology - in the hope to discard the 
mathematical apparatus of Einstein's theory in favour of the much simpler  
apparatus of Newton's theory. This is to some degree misleading, and the 
validity of the Friedman equation within Newtonian cosmology is easily 
understood: One can introduce comoving cosmological coordinates $\xi^k,\tau$ 
instead of the local PPN-coordinates $x^k,t$ used so far by requiring 

(i) that the space components of the matter 4-velocity vanish and that 

(ii) $\hat g_{00}=-1,~\hat g_{0i}=0$ in cosmological coordinates. 

\ni
 The transformation
$$ \tau = t-r^2\dot R/(2R) + o(t\epsilon^2),~ \xi^k =x^k/R + 
         o(r\epsilon) \eqno(19) $$
leads to
$$\eqalignno{ \hat g^{00} &= -1 + o(\epsilon^4), &(20) \cr
              \hat g^{0k} &= o(\epsilon^3), &(21) \cr
    \hat g^{ik} &= \delta_{ik}/R(\tau)^2 + o(\epsilon^2), &(22) \cr }$$
and satisfies (i) and (ii) to Newtonian order. It is now easy to see why the 
Newtonian scale factor $R(t)$ must be a solution of the Friedman equation: 
Consider the Friedman equation for the correct general-relativistic scale 
factor $S(\tau)$. Writing 
$$  S(\tau)=S(t-r^2\dot R/(2R)) \approx S(t) -
 \dot S r^2 \dot R/(2R) \eqno(23)$$
in this equation and expanding, terms $\sim (r/L)^2$ cancel, and one is left 
with the Friedman equation for the Newtonian scale factor $R(t)\equiv S(t)$. 

Thus it appears that the applicability of the Newtonian approximations 
(12) - (18) with the potential (10) is confined to a {\it neighbourhood 
of the observer, corresponding to distances small 
compared to the Hubble distance}. As a consequence, the matter density 
$\rho$ in local coordinates is a pure time function only on the observer 
world line $x^i=0$ and changes with $r$  further out just to {\it ensure} 
homogeneity or a Copernican principle. Indeed, any scalar depending only on 
cosmic time in cosmological coordinates, depends in a local PPN
coordinate system also on spatial coordinates (see Eq. (19)).

Many years ago, Callan, Dicke and Peebles (1962) have presented 
similar ideas concerning
the relation between Newtonian mechanics and cosmological expansion.

\stitle{3. POST-NEWTONIAN EQUATIONS}
\ni
Beyond the Newtonian approximation we use instead of (5),(6)  
the corresponding differential formulations 
($G=1$ subsequently, we also write 
$\Psi = \beta_1\Psi_1 + \beta_2\Psi_2$, assuming dust):
$$\eqalignno{\Delta \Psi &= -4\pi\rho\phi, &(24)\cr \Delta V_k 
&= -4\pi\rho v_k.&(25) \cr} $$
Differential relations also hold for the remaining functions.
$W_i$ may be written 
$$\eqalignno{ W_i &= \int d{\bf x'}\rho({\bf x'},t)({\bf x -x'}){\bf v'}
{\partial\over \partial x^{i'}}{1 \over \mid {\bf x -x'}} \mid &\cr
    & =  -x^kV^k_{,i} + {\partial \over \partial x^i} 
\int d{\bf x'}\rho({\bf x'},t) {\bf x' v'}/\mid {\bf x -x'} \mid,
& \cr} $$
or after applying the Laplacian 
$$ \Delta W_i = -4\pi\rho v_i -2 V_{k,ki}.\eqno(26)$$
Finally we list some further relationships (see Will 1993, chapter 4.1) 
already in differential form:
$$\eqalignno{ V_{k,k} &= -U_{,0}, &(27) \cr
\Delta \zeta &= -2U, &(28) \cr
\zeta_{,0k} &= V_k -W_k,&(29) \cr
\zeta_{,00} &= {\cal A} - \Psi_1.&(30) \cr} $$

The differential formulation of the PPN formalism used in this article 
is based on the (not all independent) relations (24) - (30) and (11). 
We write down expansions in terms of powers of $r$ similar to (10) 
for the post-Newtonian potentials $V_i,W_i,\Psi,{\cal A}$
as well as for the fluid density and velocity,  and look for homogeneous and
isotropic purely expanding solutions. The matter density and velocity for
isotropically moving dust can be expanded as
$$\eqalignno{\rho &=  \rho_0 + \rho_2r^2, &(31)\cr 
v^k &= x^k(f_0 + f_2r^2),&(32) \cr} $$
where $\rho_0,\rho_2,f_0,f_2$ are time functions. Using this power law 
expansion, a straightforward calculation yields

$$\eqalignno{U &= - {2\over 3} \pi \rho_0 r^2 - 
{1\over 5}\pi\rho_2 r^4, &(33)\cr 
\Psi &={1\over 15}(-3\pi\beta_1\rho_0f^2_0 +2\beta_2\pi^2\rho_0^2)r^4,&(34)\cr   
V_k &= -{2\over 5}\pi \rho_0f_0 r^2 x^k , &(35)\cr 
W_k &= -V_k, &(36) \cr
{\cal A} &= {1\over 5}\pi\rho_0(3f^2_0+ {4\over 3}\pi\rho_0) r^4. 
 &(37)\cr  }$$

Up to post-Newtonian order, the metric tensor can now be written:
$$\eqalignno{g_{00} &= -1 + br^2 +c r^4, &(38)\cr 
g_{0i} &= m r^2 x^i, &(39) \cr   
g_{ik} &= \delta_{ik}(1+gr^2). &(40)\cr }$$
The time functions $b,c,m,g$ are determined by the distribution and 
motion of matter:
$$\eqalignno{b &=  - {4\over 3} \pi \rho_0, &(41)\cr 
c &= -{2 \over 5}\pi\rho_2 + 
    {4 \over 45}\pi^2\rho_0^2( 6\beta_2 -3\zeta -10\beta)
- {1\over 5}\pi \rho_0f_0^2(4\beta_1+3\zeta), &(42) \cr   
m &=  {1\over 5}\pi \rho_0 f_0(7\Delta_1 -\Delta_2), &(43)\cr 
g &= -{4\over 3}\pi\rho_0\gamma. &(44) \cr }$$

From the conservation law of the matter tensor, $ {T^{\mu\nu}}_{;\nu} = 0$,
one obtains in Newtonian approximation, in agreement with (18),
$$\eqalignno{ \dot f_0 &= -{4\over 3}\pi\rho_0 -f_0^2, &(45)\cr
              \dot \rho_0 &= -3f_0\rho_0. &(46) }$$

The post-Newtonian terms lead to equations for the next coefficients $f_2$ 
and $\rho_2:$

$$\eqalignno{ \dot f_2 &= -4 f_0 f_2 +{2\over 15}k_1\pi \rho_0 f_0^2
 +{4\over 45}k_2 \pi^2\rho_0^2- {4\over 5}\pi \rho_2, &(47)\cr
  \dot \rho_2 &= - 5 f_0\rho_2  -5 f_2 \rho_0 
 +{2\over 3}\pi f_0\rho_0^2 (2-3\gamma), &(48) }$$ 
with 
$$\eqalignno{ k_1 &= -12\beta_1-9\zeta+42\Delta_1-6\Delta_2-20\gamma
+5, & \cr
k_2 &= -20\beta +12\beta_2 -6\zeta +21\Delta_1 -3\Delta_2 -20\gamma. &}$$

The system (45) - (48) can be integrated fairly easily: From (45), (46) 
one obtains a second-order linear differential equation for $f_0$ alone,
$\rho_0$ is determined algebraically upon solution of the $f_0$ equation. 
Similarly, if a solution of (45),(46) is known, (47),(48) can be transformed 
into a linear second-order differential equation for $f_2$, leaving an 
algebraic equation for $\rho_2$.

\stitle{4. HOMOGENEITY REQUIREMENTS}

The models obtained in this way represent 
- in a post-Newtonian approximation -  a large class of 
locally isotropic, but in general inhomogeneous expanding matter 
distributions in the neighbourhood of the observer. Restrictions for 
homogeneity may be introduced by several methods. Usually one relies   
on groups of motions. A more intuitive way is to impose homogeneity with a 
Copernican principle. The origin of the PPN coordinate system may be
associated with a freely falling observer $P$. The worldline of a second 
observer $ \bar{P}$ at rest with regard to the cosmological substratum in his 
neighbourhood is given by $x^i_0(t)$, with the three functions $x^i_0(t)$ as 
solutions of the differential equations $\dot x^i_0 = x^i_0(f_0  +f_2r^2_0).$
The coordinate transformation leading from the $P$-system 
to the $\bar{P}$-system is a {\it generalization} of the 
Chandrasekhar-Contopoulos (1967) transformation, since the second observer 
moves with acceleration. A Copernican principle should state the equivalence 
of all these systems. More specific, in the local PPN system of any comoving 
observer the expansion coefficients of matter density and cosmic velocity 
(Eqns. (31), (32)) should have the {\it same} dependence on the 
corresponding coordinate time. We may set up a Copernican principle 
for the matter density and velocity as well as for the metric or derived 
quantities such as the expansion rate of the matter congruence. It is not 
obvious whether all these principles lead to the same conclusions. 

For purpose of this note we can avoid the construction of a generalized 
Chandrasekhar-Contopoulos transformation by refering to Eq.(19):
Homogeneity for scalars $\Phi$ can be defined by a dependence on the
cosmological time $\tau$ only, hence $\Phi(\tau)=\Phi(t-r^2f_0/2) \approx
\Phi(t)- r^2f_0\dot\Phi(t)/2$, and the local expansion coefficients
$\Phi_0$ and $\Phi_2$ are related by $\Phi_2 = -f_0\dot\Phi_0/2$.
Applying this to the matter density gives with (46)
$$ \rho_2 = {3 \over 2}f_0^2\rho_0. \eqno(49)$$

It is easy to check that (49) is compatible with the differential
equations (45)-(48) if and only if the conditions 
$$\eqalignno{ 6\zeta -12\beta_1 + 20\beta +
3(\Delta_2-7\Delta_1)+26\gamma -16 & = 0, &(50)\cr
9\zeta +12\beta_1 +6(\Delta_2-7\Delta_1) +20\gamma+4 & =0. &(51)\cr
 }$$
for the PPN coefficients are satisfied. 
Moreover, $f_2$ is given by
$$f_2 = {2 \over 15}\pi\rho_0 f_0(8-3\gamma). \eqno(52)$$
(49) - (52) also ensure the homogeneity of the fluid's expansion rate up 
to post-Newtonian order via a relation similar to (49). 

The problem is that (50) and (51) hold only for a fairly small
class of gravitational theories. As one easily checks, General 
Relativity ($\zeta=0$, all other coefficients $=1$) belongs to this  class.
Needless to say that reasonable cosmological solutions are known for many 
metric theories, e.g. for the often discussed Brans-Dicke theory with 
$\gamma =(1+\omega)/(2+\omega),\beta_1 =(3+2\omega)/(4+2\omega),\beta_2 = 
(1+2\omega)/(4+2\omega), \Delta_1 = (10 +7\omega)/(14 + 7\omega)$ (other 
parameters as in GR), which do not satisfy (50) and (51). 

Obviously, the framework used so far is not adequate. To deal with 
a possible cosmological time evolution in the Newtonian dynamics of 
compact bodies, K. Nordtvedt (1993) has introduced {\it time-depending}
PPN parameters (as well as time-dependent coupling constants and
masses).
It is perhaps not necessary to go so far. Evidently the 
cosmological solution (19) of the Poisson equation is incomplete, one
could have added an arbitrary time function $U_0(t)$ to the potential. 
This also applies to the post-Newtonian potentials, giving additional
time functions $a(t),l(t),d(t)$ in the metric tensor:

$$\eqalignno{g_{00} &= -1 + a +br^2 +c r^4, &(53)\cr 
g_{0i} &=  x^i(l+ m r^2)  , &(54) \cr   
g_{ik} &= \delta_{ik}(1+ d + gr^2). &(55)\cr }$$

If the added time function vary on scales not smaller than the Hubble
time they have small or negligible influence on the motion of local bodies. 
The equations $(53)-(55)$ should allow to work out a consistent 
post-Newtonian approximation scheme, as will be discussed elsewhere. 

\stitle{5. COSMOLOGICAL COORDINATES}
\ni

To obtain different look at the problem,
we may search for solutions in cosmological coordinates, 
using symmetries and matter conservation but no other field equations. 
In a second step we transform into a local PPN coordinate system and 
compare the resulting metric with the local one following from (1)-(3) 
and (33)-(37).

Purely expanding (or contracting) dust matter  satisfies
$$\eqalignno{ T^{\mu\nu} &= \mu u^\mu u^\nu,  &(56) \cr
u_{\mu;\nu} &= \Theta(g_{\mu\nu} + u_\mu u_\nu),&(57) \cr} $$
as well as the  divergence relation
$$ {T^{\mu\nu}}_{;\nu} = 0.\eqno(58)$$

One easily verifies the existence of a comoving ($u^\mu = \delta^\mu_0$) 
coordinate system $(\tau,\xi^i)$, in which the line element can be written
$$ds^2=-d\tau^2+S(\tau,\xi^i)^2\gamma_{ik}(\xi^k)d\xi^id\xi^k.\eqno(59)$$
(58) reduces to
$$ \mu S^3 = const. \eqno(60)  $$
No field equation has been used so far. To restrict to homogeneous
models, it is assumed additionally to (56)-(58) that the scale factor 
$S$ depends on the cosmological time $\tau$ only and that 
$$\gamma_{ik}= \delta_{ik}/(1- k\xi^2/4)^2, ~~\xi^2 = \xi^k \xi^k\eqno(61)$$
is the metric of a 3-space of constant curvature. To relate the cosmological 
coordinate system to the standard PPN gauge 
up to post-Newtonian order, we extend the transformation (19) by writing
$$\eqalignno{
\tau &= u(t) +v(t)r^2 + w(t)r^4, &(62) \cr
\xi^i &=  x^i(1/F(t)  + p(t)r^2). &(63)\cr} $$

The function $u(t)$ accounts for the possibility that the local PPN time 
differs from the cosmological time even at the origin of the PPN system. 
Also, the transformation function $F$ could differ from the scale factor 
$R(t) = S(u[t])$.
A PPN coordinate system is defined  by the condition, that the quantities 
entering the metric $(1)-(3)$ must be solutions of the PPN differential 
relations. One has furthermore to ensure that up to and including the order 
$\epsilon^2$  the spatial components $g_{ik}$ attain no non-diagonal term. 
This latter condition restricts the  function $p(t)$ in (63) to 
$$ p(t) =F v^2/R^2.\eqno(64) $$

Comparing now the local and transformed metric on obtains to zero order in $r$
from the $g_{00}$ and $g_{ik}$ components
$$\eqalignno{\dot u^2  &= 1 - 2U_0 +2\beta U{_0}^2 -4\Psi_0 
+\zeta {\cal A}_0, &(65) \cr 
F &= R/\sqrt{1+2\gamma U_0}. &(66) \cr} $$

These relations cast some light on the meaning of local potentials. 
The relation between comoving and local coordinates is 
determined by the local scale factor $F(t)$, which differs from the 
cosmological scale factor $R(t)$, if a local potential $U_0(t)$
is present. 
A time dependent potential in $g_{00}$ comes not only 
from $U_0$, but also from the ${\cal A}$ and $\Psi$ terms, as seen in 
(65). If the combination of local potentials on the rhs of (65) 
differs from zero, the PPN coordinate time is not equal to the proper 
time measured by an observer on the center world line $x^k=0$ of the
PPN system (this world line is still a geodesic, however).  

In order to check the conditions for the PPN parameters found above 
we exclude all local components in $g_{\mu\nu}$. Furthermore, the relation 
(60) as well as $u^\mu = \delta^\mu_0 $ can be transformed into the
local coordinate system up to post-Newtonian order. We then compare
the transformed cosmological solution with the local metric specified by
(1)-(3) and (10), (33)-(37). Additionally  to the determination of the 
transformation functions, all these relations give the following restrictions 
for the PPN parameters:

$$ \gamma = 1,\eqno(67)$$
$$ 27(7\Delta_1-\Delta_2) -48\beta_1 +12\beta_2 -20 \beta -106 = 0.\eqno(68)$$

The domains in the PPN parameter space covered by (50), (51 on the one side 
and by (67), (68) on the other do not coincide. (Note however, if we assume 
$\gamma = 1$ in (50),(51), Eq. (68) follows from (50),(51)). But 
coincidence cannot be expected. In deriving (50), (51) we have assumed a 
restricted formulation for the Copernican principle (homogeneity for the 
matter density). Other formulations for this principle may lead to different 
conditions for the PPN parameters. On the other side, the 
discussion in this section is complete with regard to symmetries but lacks
full dynamics.

It is open if a full consideration of both symmetry conditions and dynamics 
would confine the set of theories which are free from local potentials to
General Relativity alone.

\stitle{6. CONCLUSION}
\ni
We have discussed the question, how cosmological boundary conditions 
may influence the motion of local matter. There is no influence for 
General Relativity (and a small number of additional theories of 
gravity) up to post-Newtonian order. 
For the majority of other gravitational theories there 
exists an influence at the post-Newtonian level of 
approximation. We have called this different behaviour of gravity 
theories non-Machian and Machian, respectively. In a 
post-post-Newtonian approximation the local motion of matter may 
depend on cosmological boundary conditions also in General Relativity, 
so that every metric theory of gravity would become Machian according 
to the definition given here. - Only cosmological background fields have 
been used in this article. A study of perturbations up to post-Newtonian
order may give a chance to determine some PPN parameters also in a 
cosmological setting.

\stitle{7. ACKNOWLEDGEMENTS}
\ni
I am indepted to C. M. Will and K. Nordtvedt for a useful discussion 
and to the referee for valuable suggestions concerning the presentation 
of the material.

\stitle{8. REFERENCES}
\ni

\bs
\begingroup
\parindent=0pt\everypar={\global\hangindent=20pt\hangafter=1}\par

\ref{$1.$}{Brill, D.R. (1995).}{ In: {\it Mach's Principle: 
From Newton's Bucket to Quantum Gravity}, Vol.6, {\it Einstein Studies}, 
Birkh\"auser, Boston} {} {1995}
\ref{$2.$}{Barbour, J.B. (1993).}{ In: {\it Mach's Principle: From Newton's 
Bucket to Quantum Gravity}, Vol.6, {\it Einstein Studies}, Birkh\"auser, 
Boston} {} {1995}
\ref{$3.$}{Callan, C., R.H. Dicke and P.J.E. Peebles (1965). }
{Am. J. Phys.}{33}{, 105.}
\ref{$4.$}{Chandrasekhar, S. and Contopoulos, G. (1967).}
{Proc. Roy. Soc. (London)}
{298A}{, 123.}
\ref{$5.$}{Dautcourt, G. (1990). }{Acta Phys. Pol.}{B 21}{, 755.}
\ref{$6.$}{Ehlers, J. (1981).} {In: Grundlagenprobleme der Modernen 
Physik, ed. J. Nitsch, J. Pfarr, E.W. Stachow, Mannheim} {}{1981.}
\ref{$7.$}{Heckmann, O. and Sch\"ucking, E. (1959).}{ Handbuch der 
Physik}{15}{, 489.~Berlin: Springer. }
\ref{$8.$}{Heckmann, O. and Sch\"ucking, E. (1955).}{Zeitschrift f. 
Astrophysik}{38}{, 94.}
\ref{$9.$}{Heckmann, O. and Sch\"ucking, E. (1956).}{Zeitschrift f. 
Astrophysik}{40}{, 81.}
\ref{$10.$}{Lottermoser, G. (1988).} {Ph.D. thesis, Munich.} {} {} 
\ref{$11.$}{Misner, Ch.W., Thorne, K.S. and Wheeler, J.A. (1973) 
Gravitation. San Francisco: 
W.H. Freeman and Company.}{}{}{}
\ref{$12.$}{Nordtvedt, K. (1993).}{Astrophys. J.}{407}{, 5}
\ref{$13.$}{Trautman, A. (1966).``Comparison of Newtonian and Relativistic 
Theories of Space-Time" In~}{Perspectives in Geometry and Relativity.}{}
{B. Hoffmann, ed. Bloomington and London: Indiana University Press, 
pp. 413-425.}
\ref{$14.$}{Will, C.M. (1993)  Theory and experiment in gravitational 
physics,  revised ed.   Cambridge: University Press.}{}{}{}

\endgroup
\end